\documentclass[aps,prb,showpacs,superscriptaddress,twocolumn,epsfig]{revtex4-1}
\usepackage[english]{babel}

\usepackage{graphicx}
\usepackage{amsmath}
\usepackage{hyperref}

\newcommand{\beq}{\begin{equation}}
\newcommand{\eeq}{\end{equation}}
\newcommand{\bea}{\begin{eqnarray}}
\newcommand{\eea}{\end{eqnarray}}

\newlength{\myL}

\def\be{\begin{eqnarray}}
\def\ee{\end{eqnarray}}

\begin{document}

\title {Valley-Selective  Landau-Zener Oscillations in Semi-Dirac p-n Junctions}
\author{K. Saha}
\affiliation{Department of Physics and Astronomy, University of California, Irvine, California 92697, USA}
 \affiliation{California Institute for Quantum Emulation, Santa Barbara, California 93106, USA}
 
 \author{R. Nandkishore}
\affiliation{Department of Physics, 390 UCB, University of Colorado, Boulder, CO 80309, USA}
\affiliation{Center for Theory of Quantum Matter, University of Colorado, Boulder, CO 80309, USA}

\author{S. A. Parameswaran}
\affiliation{Department of Physics and Astronomy, University of California, Irvine, California 92697, USA}

 \affiliation{California Institute for Quantum Emulation, Santa Barbara, California 93106, USA}

\date{\today}
\begin{abstract}
We study transport across p-n junctions of {gapped} two-dimensional semi-Dirac materials: nodal semimetals whose energy bands 
disperse quadratically and linearly along distinct crystal axes. The resulting electronic properties --- relevant to materials such as TiO$_2$/VO$_2$ multilayers and  $\alpha$-(BEDT-TTF)$_2$I$_3$ salts  --- continuously interpolate between those of mono- and bi-layer graphene as a function of propagation angle. We demonstrate that tunneling   across the junction depends  on the orientation of the tunnel barrier relative to the crystalline axes,  leading to strongly non-monotonic current-voltage characteristics, including negative differential conductance in some regimes.  In multi-valley  systems these features provide a natural route to engineering valley-selective transport. 
\end{abstract}
\maketitle

\section{Introduction} 
Quantum tunneling is at the heart of modern semiconductor technology:  the simplest incarnation of the venerable p-n junction relies on controlling electron tunneling primarily by tuning device parameters such as the width of the depletion region and its characteristic energy barrier. Richer behavior more sensitive to microscopic material properties is possible in the presence of multiple bands, where applied fields can induce interband Landau-Zener transitions \cite{lz}, reflected in highly nonlinear current-voltage  ($I-V$)  characteristics. Quantum interference, on the other hand, has historically been of less relevance to semiconductor devices, and was largely ignored in this context until the recent revolution in two-dimensional materials: for instance, interference phenomena stemming from the Dirac nature of electronic quasiparticles in graphene  have been successfully probed in transport experiments. Recent work \cite{rahul} involving one of the present authors examined the interplay of interference and tunneling  in bilayer graphene (BLG), whose pristine electronic dispersion consists of a pair of bands touching quadratically at isolated nodes in momentum space. As a consequence of this unusual electronic structure, when electrons in gapped BLG tunnel across a barrier between p- and n-type regions, they may be transmitted with  multiple possible phase shifts. Specifically, for a single tunnel barrier, solutions to the Schr\"odinger equation corresponding to forward propagation under the tunnel barrier come in complex conjugate pairs, whose interference causes oscillatory transmission ---  a phenomenon termed  `common-path interference'. The oscillatory tunneling probability depends on the applied bias and the gap, leading to non-monotonic current-voltage response, with negative differential conductance in specific bias regimes. While common-path tunneling interference occurs in any semiconductor \cite{sonika}, it is subleading to conventional tunneling except near quadratic band crossings --- hence its starring role in BLG, versus its relative unimportance in monolayer graphene\cite{brey} and conventional semiconductors.

This dichotomy is particularly relevant
to a new class of 2D semimetals (sometimes termed zero-gap semiconductors) with `semi-Dirac' dispersion, characterized by electronic bands meeting in a discrete set of  nodes about which the bands split linearly or quadratically along independent high-symmetry directions \cite{del, banerjee, banerjee1,gmon}. Such degeneracies can be stabilized by symmetry, and  semi-Dirac dispersion has been argued to emerge in TiO$_2$/VO$_2$ heterostructures~\cite{pardo} and $\alpha$-(BEDT-TTF)$_2$I$_3$ organic salts under pressure~\cite{kata}. There are also  proposals to engineer similar behavior photonic metamaterials~\cite{yu}, and possibilities of having such dispersion in optical lattices and graphene based materials or organo-metallic systems ~\cite{duan,dirac}. Owing to the anisotropic dispersion, semi-Dirac systems have quasiparticle properties that continuously interpolate between those of mono- and bi-layer graphene as a function of orientation. It is therefore natural to ask how this can be leveraged when building quantum devices from them --- a question that has received surprisingly little attention to date.

We study the behavior of a p-n junction in a gapped semi-Dirac material, focusing on the interband Landau-Zener problem in the presence of an applied bias. As we demonstrate below, the anisotropy of a single semi-Dirac node is imprinted on the tunneling behavior, via its dependence on the orientation of the barrier potential. When electrons propagating along the quadratic axis are incident on a barrier, they experience common-path interference and hence the tunneling amplitude oscillates with bias. The interference is absent when the barrier is oriented so that incident electrons propagate along the linear axis, with a critical angle separating the two regimes. As a consequence, the differential conductance of a p-n junction is orientation-dependent. Appropriately designed p-n junctions may therefore facilitate valley-selective transport in materials with multiple valleys with distinctly oriented semi-Dirac dispersions, potentially paving the way to several new applications.

\begin{figure}
\includegraphics*[width=0.99\linewidth]{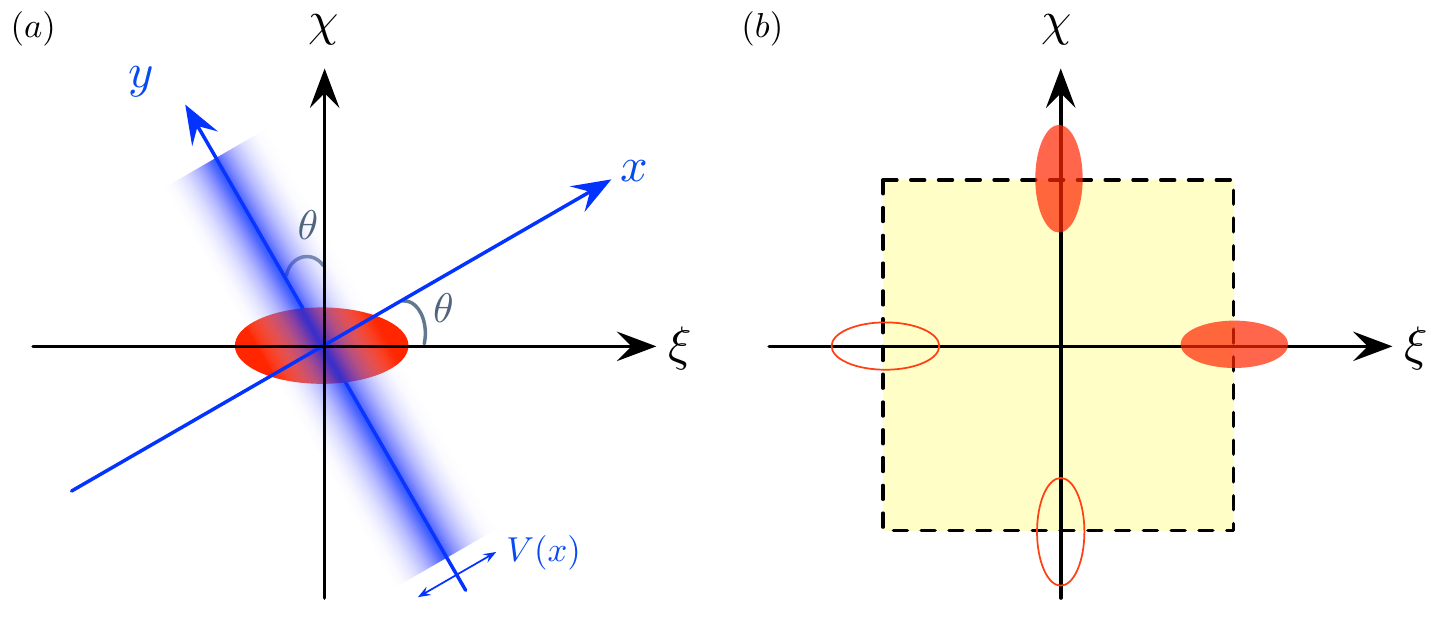}
\caption{(a) p-n junction  geometry for a single semi-Dirac node (red ellipse).  Eq.~(\ref{eq:ham0}) is defined in the $X'\equiv(\xi,\chi)$ plane with quadratic (linear) dispersion along the major (minor) axis of the ellipse. The barrier $V(x)$ is defined in the $X\equiv (x,y)$ plane oriented at an angle $\theta$ to $\xi$.  (b) Two   perpendicularly oriented nodes at the edges of the first BZ (shaded region) of a semi-Dirac material. Filled ellipses denote the nodes in the extended-zone scheme.}
\label{fig:setup}
\end{figure}
\section{Model and WKB solutions}
We begin with the low-energy Hamiltonian for a single gapped semi-Dirac node, 
\be
H=\frac{1}{2m} p_{\xi}^2\sigma_1+v p_{\chi}\sigma_2+\Delta_0\sigma_3,
\label{eq:ham0}
\ee
where we work in $\hbar=1$ units, $\sigma_i$s are Pauli matrices, $(p_{\xi},p_{\chi})\equiv -i(\frac{\partial}{\partial \xi},\frac{\partial}{\partial \chi})$ are the momenta in the $X'\equiv(\xi,\chi)$ plane ({cf. Fig.~\ref{fig:setup}a}), $m$, $v$ are the effective mass along and the Dirac velocity along  $\xi$ and $\chi$, and $\Delta_0$ is the energy gap parameter. Absent a potential, the solutions of ~(\ref{eq:ham0}) correspond to quasiparticles that disperse quadratically and linearly along $\xi$ and $\chi$ respectively.

In order to extract the dependence of Landau-Zener tunneling across a p-n junction on its orientation, we model the junction potential as a uniform field $V(x)=-F x$, in a rotated coordinate system $X\equiv(x,y) = R(\theta)X'$  where  where $R$ is a rotation about the $z$-axis. As a first pass at the problem, we attempt a solution within the Wentzel-Kramers-Brillouin (WKB) approximation. To that end, we assume $V(x)$ is slowly varying  and choose a semiclassical {\it ansatz} for the wavefunction,
\be
\psi_{\vec{p}} (\vec{x}) \sim \left(\begin{array}{c} \psi_1 \\\psi_2 \end{array}\right)e^{ip_y y}\exp\left[{ i\int_{-\infty}^{{x} } {p}_x({x'})\,  dx'}\right]
\ee
where  ${p}_x(x)$ is a slowly varying function of $x$ and we have used translational invariance in $y$ to fix the wavefunction to be a plane wave parallel to the barrier. Using this in (\ref{eq:ham0})  we find that for a solution at energy $E$, $p_x(x)$ satisfies
\be
(E-V(x))^2 &=&(2m)^{-2} (p_{x}(x)\cos\theta -p_{y}\sin\theta )^4\nonumber\\
& &+v^2 (p_{x}(x)\sin\theta +p_{y}\cos\theta )^2+\Delta_0^2.
\label{eq:energy}
\ee
In order to illustrate the angular dependence, let us  choose normal incidence ($p_y=0$) for simplicity and work at threshold ($E=0$). {Then the solution of Eq.~(\ref{eq:energy}) reads 
\be
p_{x}(x)=\pm \frac{\sqrt{2}m v\sin\theta}{\cos^2\theta}\left[-1\pm i\sqrt{\frac{\cot^4\theta}{m^2v^4}(\Delta^2_0- V(x)^2)-1}\right]^{1/2}.\nonumber\\
\label{eq:momn}
\ee
This can be further simplified by writing the part within braces of Eq.~(\ref{eq:momn}) as $(\pm i p_0+p_1)^2$, where
\be
p_0(x)=\frac{\sqrt{1+\frac{\cot^2\theta}{mv^2}\sqrt{\Delta^2_0- V(x)^2}}}{\sqrt{2}}\nonumber\\
p_1(x)=\frac{\sqrt{-1+\frac{\cot^2\theta}{mv^2}\sqrt{\Delta^2_0- V(x)^2}}}{\sqrt{2}}.
\label{eq:papb}
\ee
With this, we find the effective WKB action $S(\theta)=\int p_x(x)dx$} in the barrier region  defined by $|V(x)|<\Delta_0$ varies with the orientation of the barrier potential. Specifically, the four solutions {of Eq.~(\ref{eq:momn}) lead to WKB actions of the form $S(\theta)=\pm i S_0 \pm S_1$, where the signs are chosen independently and
\be
{S_{0,1}}= \delta\frac{\sin\theta}{\cos^2\theta}\int_{-1}^{1} d\tilde{x}  \left[\pm1 +  {{\eta}\delta} \cot^2\theta\sqrt{1-\tilde{x}^2}\right]^{1/2},
\label{mon1}
\ee
where $\tilde{x} = F x /\Delta_0$, $\delta = \Delta_0/\Delta$ and $\eta = F/m^2 v^3$ are dimensionless, $\lambda = 1/mv$ is an intrinsic length scale of the material, and $\Delta =  F\lambda$ is the characteristic energy scale of the junction.
 $S_0$ is real and positive for all $\theta$, but  $S_1$ sensitively depends on the orientation of the barrier. 
  Above a bias-independent critical angle $\theta_c= \cot^{-1} (v/\sqrt{2\alpha\Delta_0})=\cot^{-1}(1/\sqrt{\eta \delta})$,  $S_1$ becomes purely imaginary, so that all four WKB solutions correspond to monotonically decaying exponentials, as in conventional p-n junctions. 
In contrast, for $\theta<\theta_c$, $S_1$ has a non-zero real part,  $S_1=S_1'+i S_1''$, where $S'$ and $S''$ are real and positive. The WKB solutions then take the form $S(\theta)=\pm \left[i (S_0+\zeta S_1'')+ \zeta S_1'\right]$ where $\zeta = \pm 1$; since we are interested in physical solutions that decay into the barrier, we focus on those with ${\rm Im}(S)>0$, 
leading to a WKB wavefunction in the barrier region of the form
\be
\Psi=A e^{- (S_0 + S_1'')+iS_1'}+A^{\ast} e^{-(S_0-S_1'')-iS_1'},
\ee  
where $A=|A| e^{i\varphi}$ is a complex number, and we have used the fact that $S_0 \pm S_1'' >0$.
\begin{table}[t]
\begin{tabular}{ |c|c|c|c|c|c| }
\hline
 Material & $m/m_e$ & $v$ (m/s)&$\lambda (\AA)$ &$\Delta_0$ (eV)&$\theta_c$  \\ 
 \hline
 (TiO$_2)_5$/(VO$_2)_3$			&$13.6$ 	& $1.5\times10^5$	&$0.56$ 	&$0.5$ 	&$0.16\pi$	 \\  
$\alpha$-(BEDT-TTF)$_2$I$_3$	&$3.1$ 	& $1.14\times10^5$	&$3.27$ 	&$0.1$ 	&$0.19\pi$ 	\\
Photonic crystals			 	&$1.2\times 10^{-3}$ &$9.9\times10^6$	&$96.5$	&$1.0$	&$0.28\pi$ 	\\ 
 \hline
\end{tabular}
\caption{\label{tab:params}Microscopic parameters for semi-Dirac materials, with representative gaps and corresponding critical angles. $m_e$ denotes the free electron mass\cite{pardo,param,yu}.}
\end{table}

The WKB phase shift $\varphi$ may be determined analytically by examining classical turning points; we instead obtain it  numerically (see below), and find $\varphi \approx \pi/2$. The
 total transmission is given by 
\be
T(\theta) =|\Psi|^2=2 |A|^2e^{-2 S_0}\left[\cosh 2S_1''+\cos 2 (S_1'-\varphi)\right].\,\,\,\,\,\,
\label{tran}
\ee
Since $S_0$ and $S_1$ are monotonic functions of bandgap and field strength, the transmission through the p-n junction oscillates as these are varied (Fig.~\ref{fig:transmission}(a)).   The oscillations resemble those in BLG~\cite{rahul}, but differ in that the nodes of transmission are now angle-dependent, occurring when $\text{Re}\,S_1-\varphi=n\pi/2$ with $n$ an odd integer. 

So far, we have focused on $p_y=0$; we now comment on WKB solutions for finite $p_y$. For $p_y\ne 0$, we obtain four complex solutions for all values of $\theta$. As before, two of these solutions correspond to waves decaying into the barrier; focusing  on these, we obtain WKB amplitudes with different decay lengths and different phase factors, in contrast to the $p_y=0$ case.  For small $p_y$ and $\theta\ll\theta_c$, these differences in decay lengths and phases are small, and we still see oscillations in the transmission (cf. Appendix~\ref{ap:transobq}). 
However for larger angles and transverse momenta, the differences  are sufficiently appreciable that the interference is much less effective due to the mismatch in amplitudes. A knowledge of the full dependence on $p_y$ is essential in order to compute $I-V$ characteristics of the junction; we determine this numerically (cf. Appendix~\ref{ap:transobq}). Table~\ref{tab:params} provides estimates of parameters for some semi-Dirac systems and the corresponding critical angles.
\begin{figure}
\includegraphics[width=0.99\linewidth]{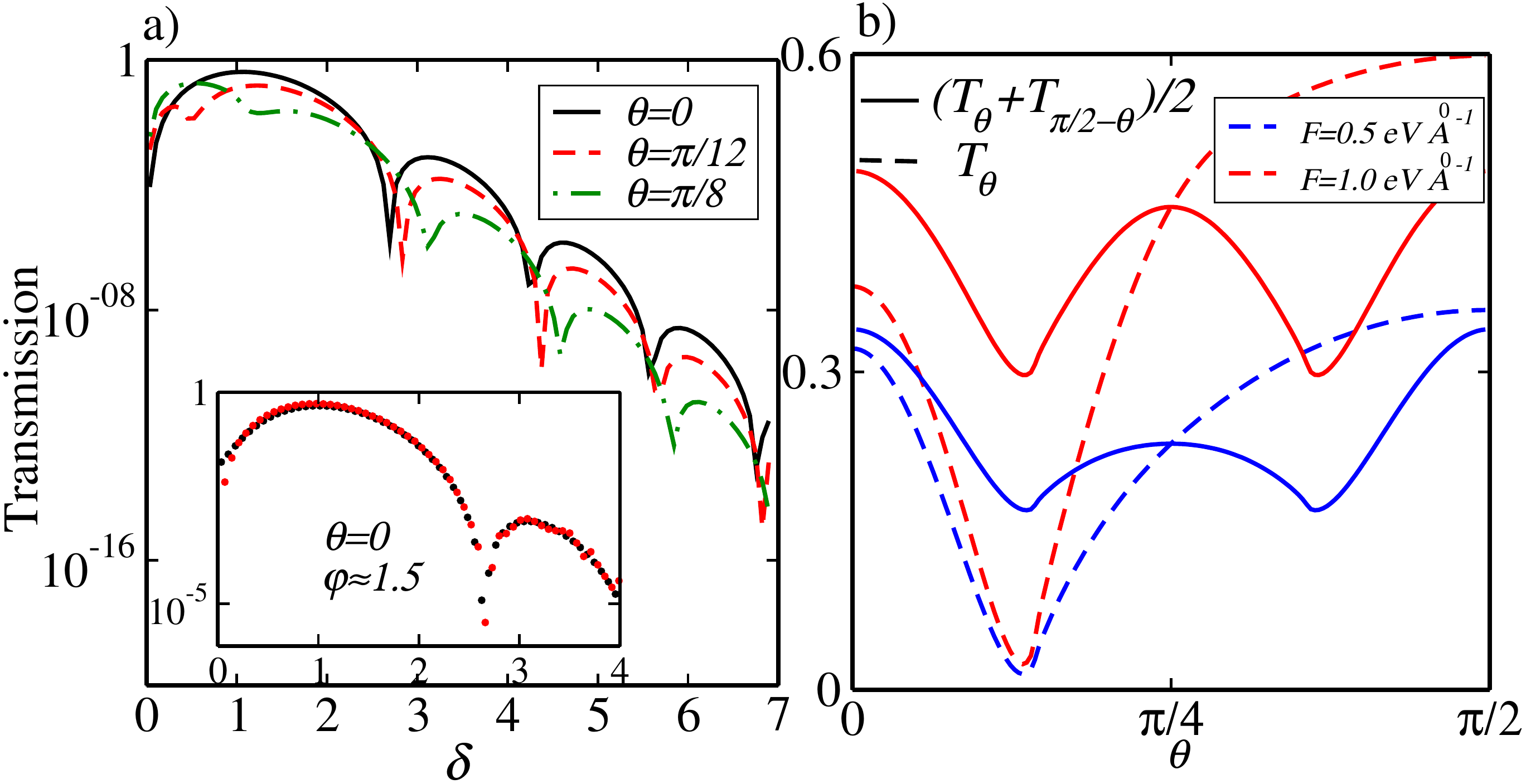}
\caption{(a) Transmission at normal incidence as a function of dimensionless gap parameter $\delta$ for different barrier angles $\theta$. (Inset)  Comparison between numerical and WKB results at $\theta=0$; a fit yields WKB phase shift $\varphi\simeq 1.5$. (b) Transmission for semi-Dirac systems with a single node(dashed line) and with two perpendicular nodes (solid line), as a function of barrier angle $\theta$. At a critical angle $\theta_c$, common-path interference gives way to conventional tunneling, for a single node, reflected by a dip in transmission. For two nodes, common-path interference is absent for $\theta_c <\theta<\pi/2-\theta_c$.  
We take $\theta_c<\pi/4$, achievable with $\Delta_0=0.5$ eV in (TiO$_2$)$_5$/(VO$_2$)$_3$.}
\label{fig:transmission}
\end{figure}

\subsection{Mapping to Landau-Zener Problem} An alternative approach (restricted to the uniform-field model) uses the fact that  
(\ref{eq:ham0}) 
 can be mapped into a Landau-Zener problem by working in momentum space, where $V(x) \rightarrow i F \partial_{p_{x} }$. We then find the first-order equation
 \be
iF{\partial\psi\over\partial p_{x}}=(H(p_x,p_y)-E)\psi,
\label{eq:schr}
\ee 
that describes the unitary evolution of a particle with $p_x$ playing the role of ``time''. The Zener tunneling probability is the square of the amplitude for an eigenstate of $H$ at $p_x=-\infty$ with negative energy to evolve to one at $p_x=\infty$ with positive energy.  It is therefore  convenient to work in the basis of eigenstates of  (\ref{eq:ham0}) for $p_x = \pm\infty$, given by $\varphi_{\mp}=(1,\mp1)^T$ where $T$ is the matrix transpose. Substituting $\psi=a\varphi_-+b\varphi_+$ (with $|a|^2+ |b|^2=1$) into (\ref{eq:schr}), we find a set of coupled differential equations for the complex amplitudes $a,b$. Integrating these equations (cf. Appendix~\ref{ap:lznm}) with the initial condition $a(p_x\rightarrow-\infty) = 1$, we find $b(p_x\rightarrow\infty)$ and hence the transmission probability. Fig.~\ref{fig:transmission}(a, inset) shows a representative comparison between WKB and numerical results for a small range of $\delta$, and for $p_y, E=0$, indicating good agreement. By fitting the parameters of the WKB and numerical results, we find $\varphi \approx1.5 \approx \pi/2$. The numerical results are unstable for large $\delta$ and for some higher $p_y$ values, but as their primary utility is in computing the WKB phase shift, this does not pose a serious problem. In the remainder, we use WKB results with $\varphi \approx \pi/2$ unless otherwise stated.
\begin{figure}
\includegraphics*[width=0.94\linewidth]{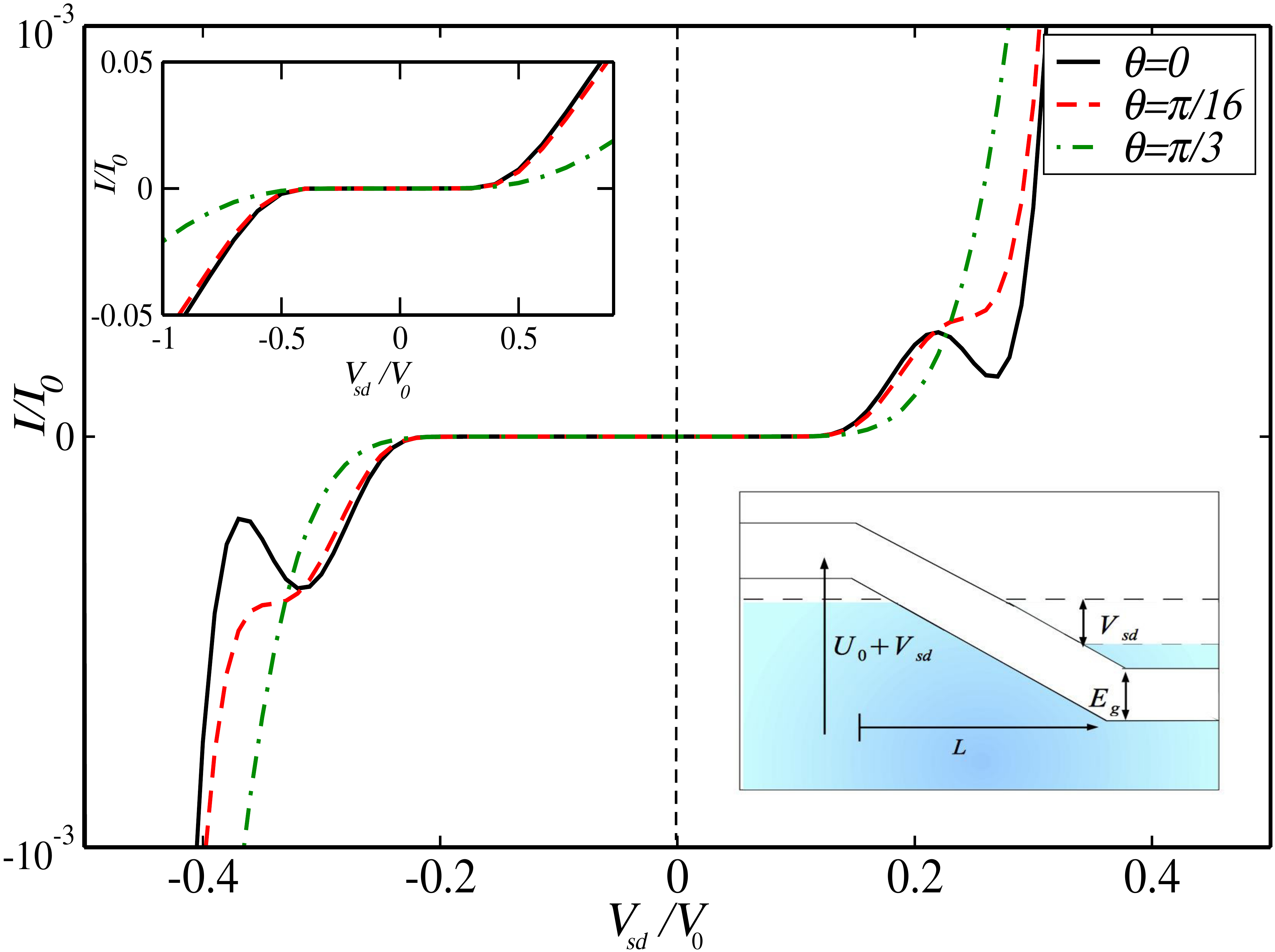}
\caption{Tunneling current $(I/I_0)$ through the p-n junction with energy gap $E_g = 2\Delta_0$ (shown schematically, bottom inset) as a function of bias voltage $V_{\text{sd}}/V_0$ for different values of $\theta$, clearly showing N-shaped features and negative differential conductivity. 
We take $\lambda$ appropriate to (TiO$_2$)$_5$/(VO$_2$)$_3$, $\Delta_0=2$ eV and $U_0/V_0=0.05$. Units are  $I_0=g\frac{e^2}{h}(\frac{W}{\lambda})V_0$ and  $V_0=\Delta_0 (L/\lambda)$.  (Top inset) Zoomed-out view of $I$-$V$ curve. } 
\label{fig:tunnelcur} 
\end{figure}

\section{I-V Characteristics} Within the Landauer formalism, the contribution of a single node to the net tunneling current across a junction of lateral width $W$ and orientation $\theta$ with applied voltage $V_{\text{sd}}$ is given by
\be
 I&=&\frac{e}{h}\int_{-\infty}^{\infty} d\epsilon \left[f_{\epsilon-\frac{1}{2}eV_{\text{sd}}}-f_{\epsilon+\frac{1}{2}eV_{\text{sd}}}\right]T_\epsilon(\theta,F),\nonumber\\
  T_\epsilon(\theta,F)&=& \frac{g W}{2\pi} \int_{-\infty}^{\infty} T(p_y,\theta,F) dp_y,
 \label{eq:gencur}
 \ee
where $f_x=(\exp(x/k_B T)+1)^{-1}
$ is the Fermi-Dirac distribution at temperature $T$, and $g$ counts the spin degeneracy. We have used the fact that the transmission is dominated by small-$p_y$ values to ignore the dependence of the Fermi occupation factors on momentum. Ignoring the energy-dependence of $T$ (valid in the uniform-field model) and writing the effective field in the barrier region as $F = \frac{e}{L}\left(U_0 + V_{\text{sd}}\right)$, where $U_0/L$ is the `built in' gate-induced electrostatic potential difference across the junction  and $L$ is the size of the depletion region, the tunneling current at low temperatures is
\be
 I \approx \frac{e^2}{h} V_\text{sd} \left[ g\frac{W}{2\pi}  \int_{-\infty}^{\infty}T\left(p_y,\theta,\frac{e}{L}(U_0 + V{_\text{sd}})\right) dp_y\label{eq:tunncur1}
 \right].
\ee

It is convenient to define  $I_0=g\frac{e^2}{h}(\frac{W}{\lambda})V_0$,  $V_0=\Delta_0 (L/\lambda)$, as natural units for the current and bias voltage. Fig.~\ref{fig:tunnelcur} shows the results of numerical integration of (\ref{eq:tunncur1}) for a reasonable value of the built-in potential, $U_0/V_0 \approx 0.05$. Note the N-shaped features in the tunnel current, with negative differential conductivity similar to an Esaki diode~\cite{esaki} (though with a different origin), as well as Zener-like threshold behavior where the current rises sharply above a breakdown voltage. These features were previously predicted for BLG p-n junctions with the crucial difference here being the sensitive angular dependence:  the  $N$-shaped features and negative $dI/dV$ are absent when the junction angle exceeds the critical $\theta_c$ for common-path interference. This angular dependence is particularly salient to devices built from multi-valley semi-Dirac materials, to which we now turn.


\begin{figure}
\includegraphics[width=0.95\linewidth]{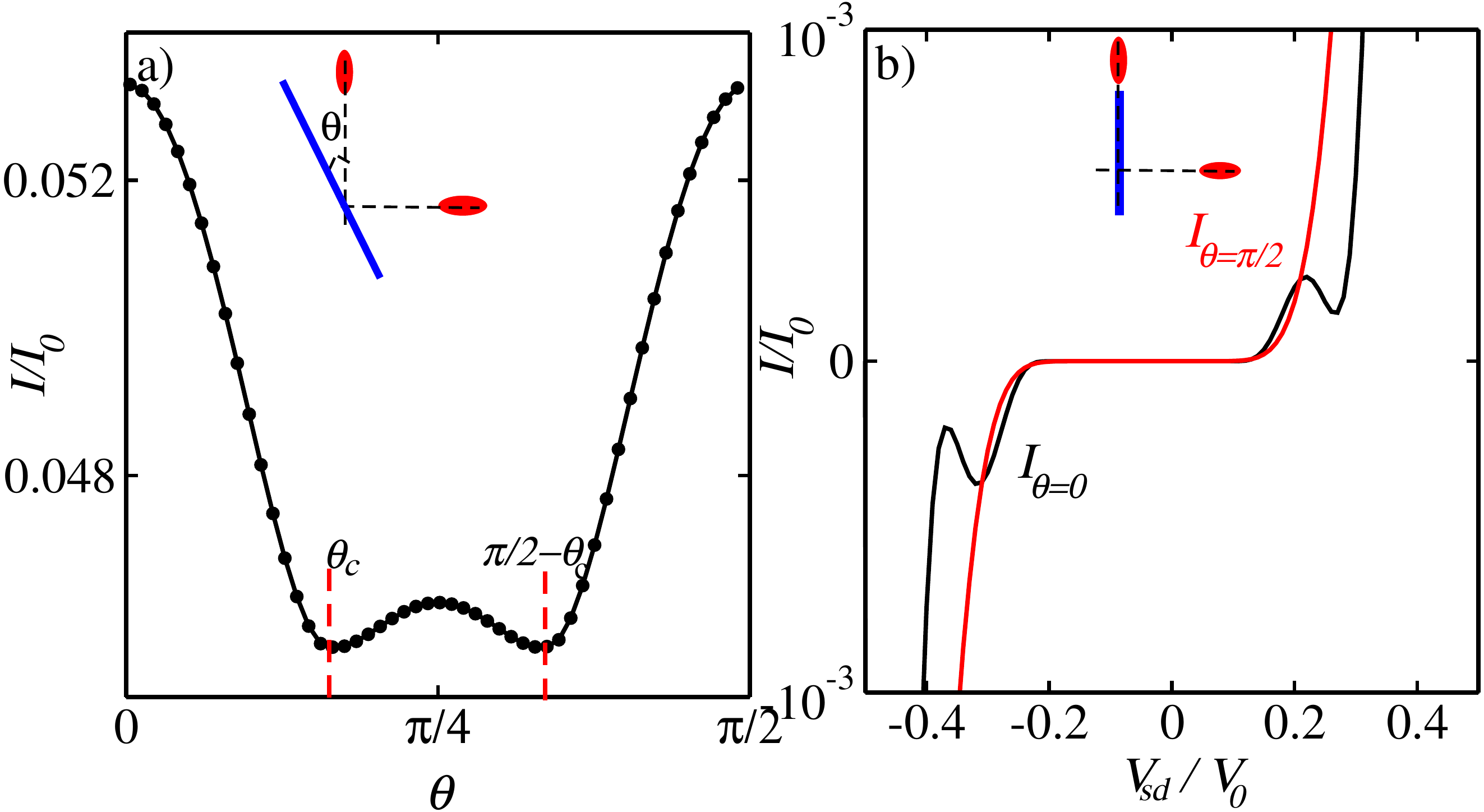}
\caption{(a) Tunneling current in a two-valley system and its  dependence on barrier angle $\theta$, for fixed bias $V_{\text{sd}}/V_0=0.5$ and  parameters are appropriate to (TiO$_2$)$_5$/(VO$_2$)$_3$ [Table~\ref{tab:params}]. For $\theta<\theta_c$ ($\theta>\pi/2-\theta_c$) the current is dominated by common-path interference contributions from node $A$ (node $B$); these oscillate with applied bias.  For $\theta<\theta<\pi/2-\theta_c$, the tunneling is solely due to conventional tunneling and does not oscillate with bias. (b) Tunneling current contributions of nodes $A, B$  as a function of bias, for a barrier whose normal is parallel to the quadratic axis of node $A$. As only the node $A$ current (black curve) oscillates with bias, the current is valley-filtered (i.e., dominated by node $B$ current, red curve) when tunneling electrons in node $A$ experience destructive common-path interference. The small residual node $A$ current at the minima is due to electrons at non-normal incidence. As in Fig.~\ref{fig:tunnelcur}, we choose a slightly larger gap $\Delta_0=2$eV  for illustrative reasons.
}
\label{fig:multivalley}
\end{figure}
\section{Multiple Valleys} In order to facilitate a more direct comparison with experimental systems, and to study valley-selective transport it is useful to discuss situations with multiple semi-Dirac nodes (`valleys') in the Brillouin zone (BZ). As minimal model consider a pair of valleys, denoted $A, B$ with quadratic axes rotated by 90$^{\circ}$ relative to each other {(cf. Fig.~(\ref{fig:setup})b)}; the generalization to multiple such pairs, or to different relative orientations, is straightforward. We take the tunnel barrier to be oriented at an arbitrary angle $\theta$ with respect to the quadratic axis  of  valley A, and hence at $\pi/2 - \theta$ with respect to that of valley $B$. The transmission through the junction is then given by summing contributions from the two valleys, viz.:
\be\label{eq:tottran}
T_{\text{tot}}(\theta) = T(\theta)+ T\left(\frac\pi2 - \theta\right).
\ee
Recall that the dominant contribution to common-path interference is from electrons incident normally on the barrier (i.e., with $p_y =0$). In this limit, the presence (absence) of this effect is sensitive to whether $\theta <\theta_c$ ($\theta > \theta_c$): for $\theta_c >\pi/4$, (\ref{eq:tottran}) always includes a common-path interference contribution from at least one node for any $\theta$, whereas if $\theta_c <\pi/4$ common-path interference is absent for $\theta_c< \theta < \frac{\pi}{2} - \theta_c$. As a consequence, in many cases the tunnel current is strongly suppressed in this window (Fig.~\ref{fig:multivalley}a). Since $\theta_c$ depends sensitively on material parameters,  it may be possible to achieve  `switching' behavior of the tunnel current e.g. by tuning $\Delta_0$. Furthermore, if we choose the barrier angles $\theta=0,\pi/2$, then only one of the nodes experiences common-path interference and hence has a complex $I$-$V$ characteristic; therefore, for an appropriate choice of bias the current through the junction is valley-filtered, i.e. dominated by one of the nodes (Fig.~\ref{fig:multivalley}b).  This quite remarkable property of the semi-Dirac system enables valley-selective functionality with a relatively conventional device design. We have focused here on the two-valley example; in the case of TiO$_2$/VO$_2$ multilayers, there are actually four nodes, one pair of which is oriented at 90$^{\circ}$ to the other pair. In such situations, the valley-selective behavior of the junction is restricted to producing a current dominated by electrons from one pair of nodes or the other. 

{To this end, we briefly comment on the experimental realization of such phenomena. Semi-Dirac p-n junctions can be constructed in a two-gate geometry similar to that in graphene\cite{wilms}, and a combination of local top gating and global back gating permits control of control carrier type and density near the barrier regime.  Moreover, the gap  in this system can be controlled by charging these two gates with opposite parity\cite{blg},  in turn allowing the tuning of  $\theta_c$ to realize angle-dependent tunneling. Since  for a typical p-n junction the band gap can be tuned by up to a few hundreds of an meV, different values of $\theta_c$ as shown in Table I may be achieved to probe the angle-dependent oscillatory transmission and tunneling current discussed here. The gate-induced `built-in' potential $U$ may be tuned by doping or via a third gate, permitting access to a reasonable number of nodes in the transmission probability. For $\Delta_0=500 meV$ gap, barrier width $L\sim 10 \AA$ and 'in-built' potential $U=2\Delta_0$ and $\lambda~0.5 \AA$, we obtain $\delta=\Delta_0/F\lambda\sim 10$, where we have used $F=U/L$.} 

\section{Concluding Remarks} We have studied the tunneling across barriers between p- and n-doped regions of semi-Dirac materials, and examined the impact of their anisotropic dispersion on transport through the junction. We find, first, that the role of common-path interference in the Landau-Zener tunneling problem is sensitively dependent on the angle of the tunneling barrier relative to the quadratic axis of the material:  for instance, electrons incident normally on the barrier do not experience interference once the barrier angle exceeds a critical value $\theta_c$. This is in turn reflected in the behavior of the $I-V$ characteristic, and specifically in the presence or absence of bias regimes with negative differential conductance. Our analysis was twofold: first, we used a WKB treatment of tunneling under the barrier that can be readily generalized to more complicated barrier potential profiles; second, we performed a Landau-Zener analysis, restricted to the case of a uniform barrier field, that served as a check on the WKB results. We also generalized these ideas to the multi-valley setting, where we propose  that the angular dependence of the Landau-Zener tunneling may be leveraged to build a ``valley valve'', as we find bias regimes where the current flowing across the junction is  valley-filtered. Although the basic ingredients of the devices proposed here are quite conventional,  the semi-Dirac system remains relatively unexplored from a device engineering perspective, so there may be unforeseen difficulties in building good p-n junctions and controlling their properties. However, we note that our analysis is quite general and applies to any material with semi-Dirac dispersion;  individual materials may have widely differing critical angles and bias regimes for the effects considered here, making them more or less suitable for specific applications. Farther afield, we note that there are also proposals to engineer semi-Dirac dispersion in photonic crystals; in such situations, our analysis suggests a promising new route to  controlling the flow of light through photonic devices.

\acknowledgements 
We thank W. Pickett and I. Krivorotov for useful discussions. We acknowledge support from NSF grant DMR-1455366 and a UC President's Research Catalyst Award CA-15-327861 (SAP, KS).

\appendix
\section{Transmission for oblique incidence}\label{ap:transobq}
In this appendix, we focus on the tunneling amplitudes at finite $p_y$. We rewrite Eq.~(3) of the main text as
\begin{figure}[t]
\includegraphics[width=0.9\linewidth]{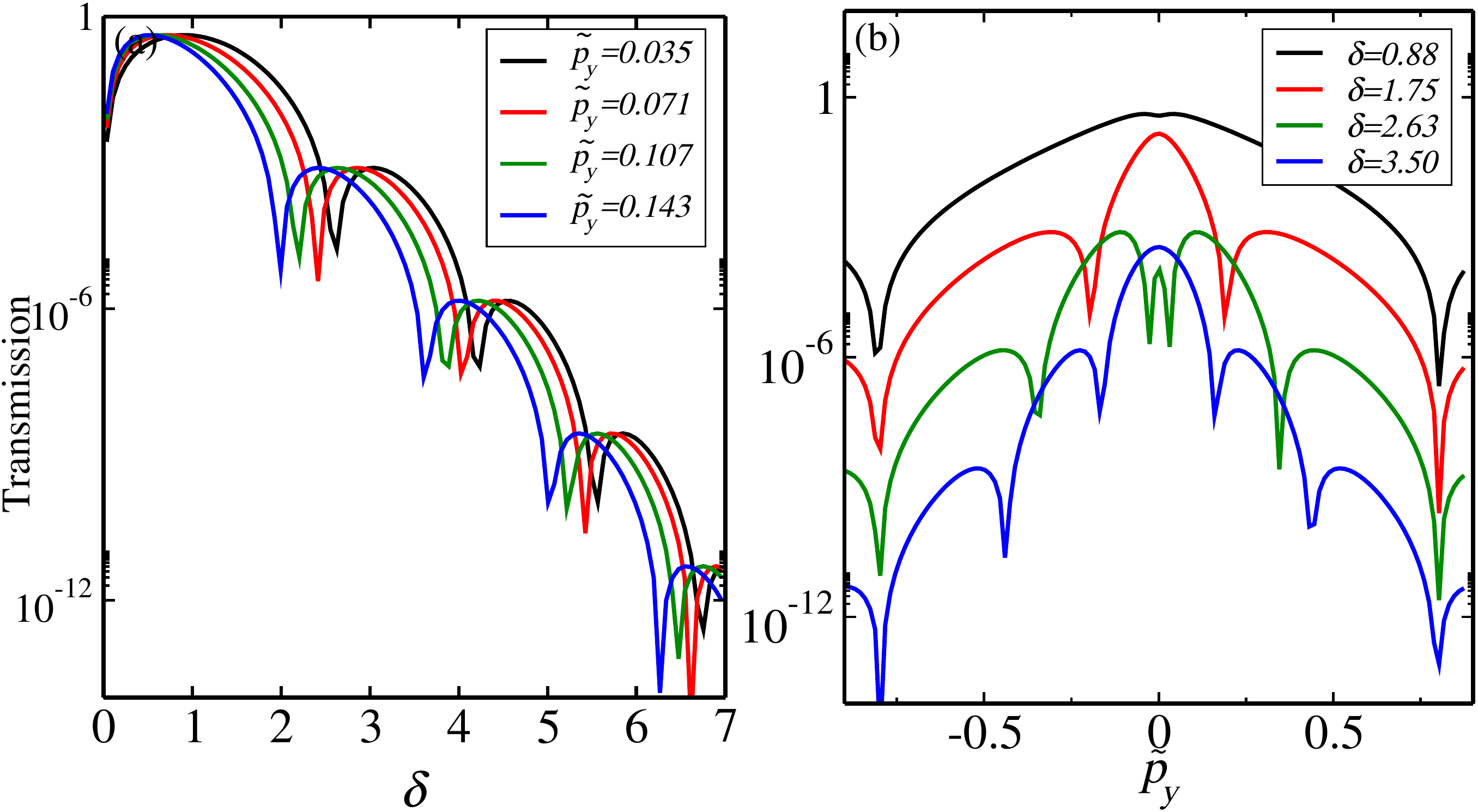}
\caption{a) Transmission as a function of $\delta$ for different values of $\tilde p_y$, where $\tilde p_y=\lambda p_y/2$. b) Same as (a) as a function of $\tilde p_y$ for different values of $\delta$. We take $\theta=0$ for both (a) and (b).  Note that, $\eta$ is kept fixed throughout this calculation. We take parameters appropriate to (TiO$_2)_5$/(VO$_2)_3$. }
\label{ky:trandel}
\end{figure}

\begin{align}
\alpha \tilde p_x(\tilde x)^4+\beta \tilde p_x(\tilde x)^3+\gamma \tilde p_x(\tilde x)^2+\zeta \tilde p_x(\tilde x)\nonumber\\
~~~~~~~+m_0+(1-V(\tilde x)^2)=0,
\label{eq:sol}
\end{align}
where $\alpha=\frac{\cos^4\theta}{4\delta^6\eta^2}, \beta=- \frac{2\tilde p_y \cos^3\theta \sin\theta}{\delta^5\eta^2},\gamma=\frac{ \sin^2\theta+6 \tilde p_y^2 \cos^2\theta \sin^2\theta}{\delta^4\eta^2}, \zeta= 4\frac{ \tilde p_y \sin\theta \cos\theta-2 \tilde p_y^3 \cos\theta\sin^3\theta}{\delta^3\eta^2}$ and $m_0=\frac{4 (\tilde p_y^4 \sin^4\theta+ \tilde p_y^2 \cos^2\theta)}{\delta^2\eta^2}$; $\eta=F/m^2v^3$ and $\delta=\Delta_0/\Delta$, while $\Delta=F\lambda$ with $\lambda=1/mv$; $\tilde p_y=\lambda p_y/2$, $\tilde p_x=\Delta_0 p_x/F$, and $V(\tilde x)=\tilde x$, where $\tilde x=F x/\Delta_0$. Note that we have written Eq.~(\ref{eq:sol}) in dimensionless units.

In the barrier regime $|V(\tilde x)|<1$,  the effective WKB action $S=\int \tilde p_x(\tilde x) d\tilde x$ leads to four solutions:  
\begin{align}
S_{i}=S'_i+i S''_i,
\end{align} 
where $i\in \{1,2,3,4\}$, and all $S'$ and $S''$ are real numbers. Of these four solutions, two propagate to the right and two to the left. Focusing on the right-moving solutions, we obtain WKB amplitudes as discussed in the main text. For $\theta=0$, the decay lengths and the phase factors have identical  magnitude for all $i$. Thus the tunneling probability oscillates, irrespective of the value of $\tilde p_y$. This is illustrated in Fig.~(\ref{ky:trandel})a for different values of $\tilde p_y$, for $\theta=0$.  Fig.~(\ref{ky:trandel})b, shows transmission as a function of $\tilde p_y$ for fixed $\delta$. Note that for fixed $\delta$, the transmission also decays as $\tilde p_y$ increases. Thus large $\tilde p_y$ solutions contribute insignificantly to the integrated transmission and hence to the net current.

On the other hand, for $\theta\ne 0$ and $\tilde p_y\ne0$, both the decay lengths and the phase factors differ in magnitude. Consequently, the oscillatory features of the transmission slowly die off with the increase of both $\theta$ and $\tilde p_y$. This is illustrated in Fig.~(\ref{del:tranky}). For small $\tilde p_y$, we see oscillatory transmission as long as $\theta\ll\theta_c$. In contrast, for $\theta>\theta_c$, we see more conventional decaying transmission with no oscillations. For large $\tilde p_y$, the tunneling amplitudes are almost decaying for both limiting cases of $\theta$. Thus for large $\theta$ and large $\tilde p_y$ we obtain conventional transmission, and hence no N-shaped feature in the integrated current.

\begin{figure}[t]
\includegraphics[width=0.99\linewidth]{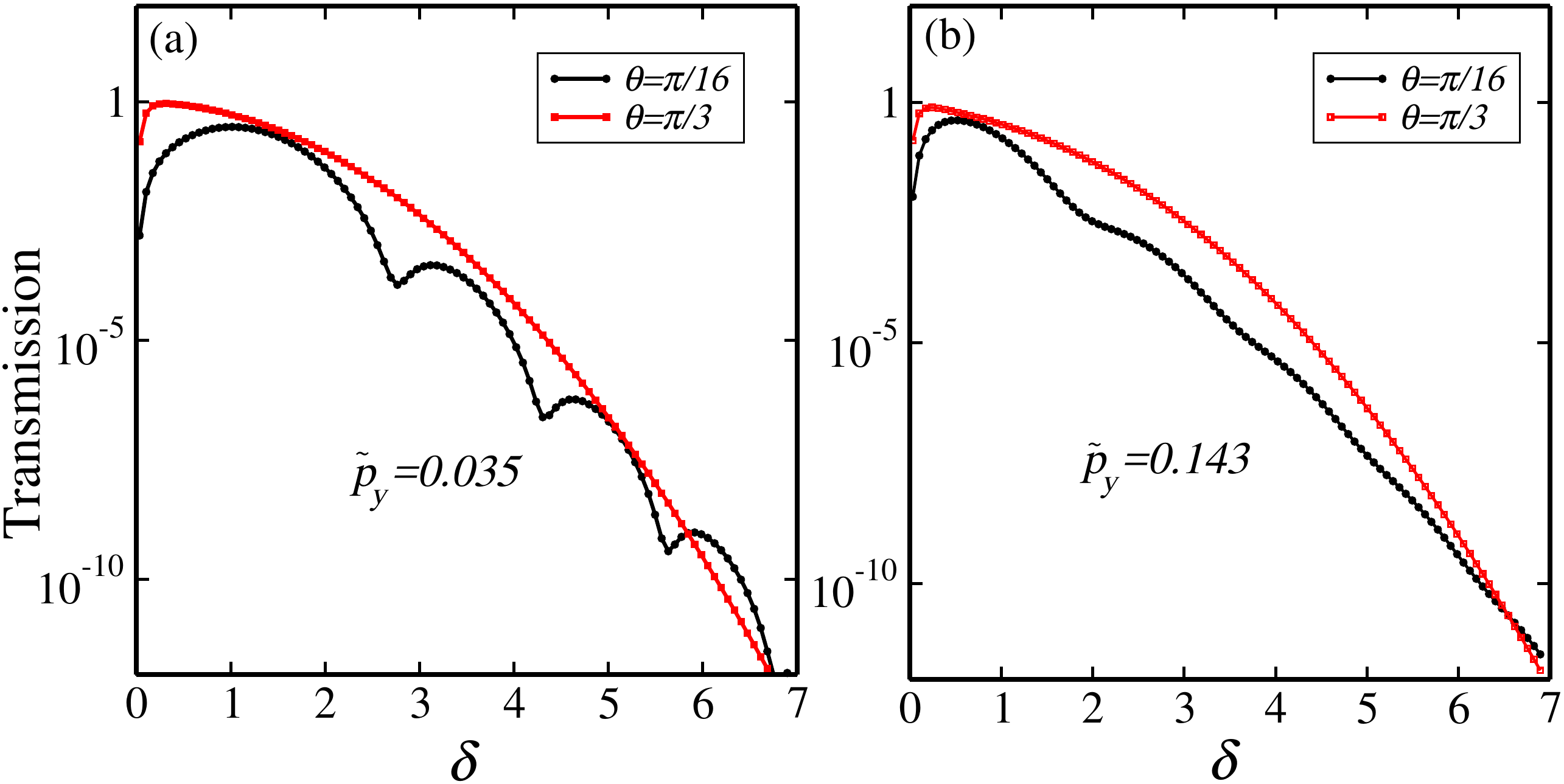}
\caption{ a) Transmission at finite $\tilde p_y$ and finite $\theta$. For small $\tilde p_y$ and $\theta\ll\theta_c$, the oscillatory feature of the transmission persists (black line), however the minima do not extend to zero in contrast to $\tilde p_y=0$ case. For $\theta>\theta_c$,  we do not see any oscillation in transmission (solid red line). This is attributed to the fact that the phase factors of right moving solutions differ significantly for $\theta>\theta_c$. b) Same as (a) for different values of $\theta$ while  $\tilde p_y$ is large. This clearly tells us that the oscillatory transmission is lost if either of $\theta$ or $\tilde p_y$ becomes large. Parameters are taken same as in Fig.~(\ref{ky:trandel}). }
\label{del:tranky}
\end{figure}

\section{Numerical solutions of Landau-Zener problem}\label{ap:lznm}
In this appendix, we provide a detailed discussion of the numerical solutions of tunneling amplitudes presented in the main text. Writing Eq.~(1) of the main text in rotated coordinate and mapping it into a Landau-Zener problem, we obtain first-order differential equation in momentum space as 
\begin{align}
iF\partial_{p_x}\psi =\left[(2m)^{-1}(p_{x}\cos\theta-p_{y}\sin\theta)^2\sigma_x+v (p_x\sin\theta\right. \nonumber\\
\left.+ p_y\cos\theta)\sigma_y+\Delta_0\sigma_z-E\right]\psi.
\label{eq:diff}
\end{align}
Note that we take $V(x)=-iF\partial_{p_x}$. Following the definition of Zener tunneling amplitudes, it is convenient to work in the eigenstate basis of $H$ at $p_x=\mp\infty$, which are given by $\varphi_{\mp}=(1,\mp1)^T$, where $T$ is  matrix transpose. Substituting $\psi=a\phi_{-}+b\phi_{+}$ (with $|a|^2+|b|^2=1$) into Eq.~(\ref{eq:diff}) and setting $E=0$ for simplicity, we obtain (in dimensionless units)
\begin{align}
i  \partial_{\tilde p_{x}}a=(4/\eta)(\tilde p_{x}\cos\theta-\tilde p_{y}\sin\theta)^2a \nonumber\\
-i (4/\eta)(\tilde p_x\sin\theta +\tilde p_y\cos\theta)a+\delta b\nonumber\\
i \partial_{\tilde p_{x}}b=-(4/\eta)(\tilde p_{x}\cos\theta-\tilde p_{y}\sin\theta)^2b \nonumber\\
+i (4/\eta) (\tilde p_x\sin\theta +\tilde p_y\cos\theta)a+\delta a.
\end{align}

\begin{figure}[t]
\includegraphics[width=0.98\linewidth]{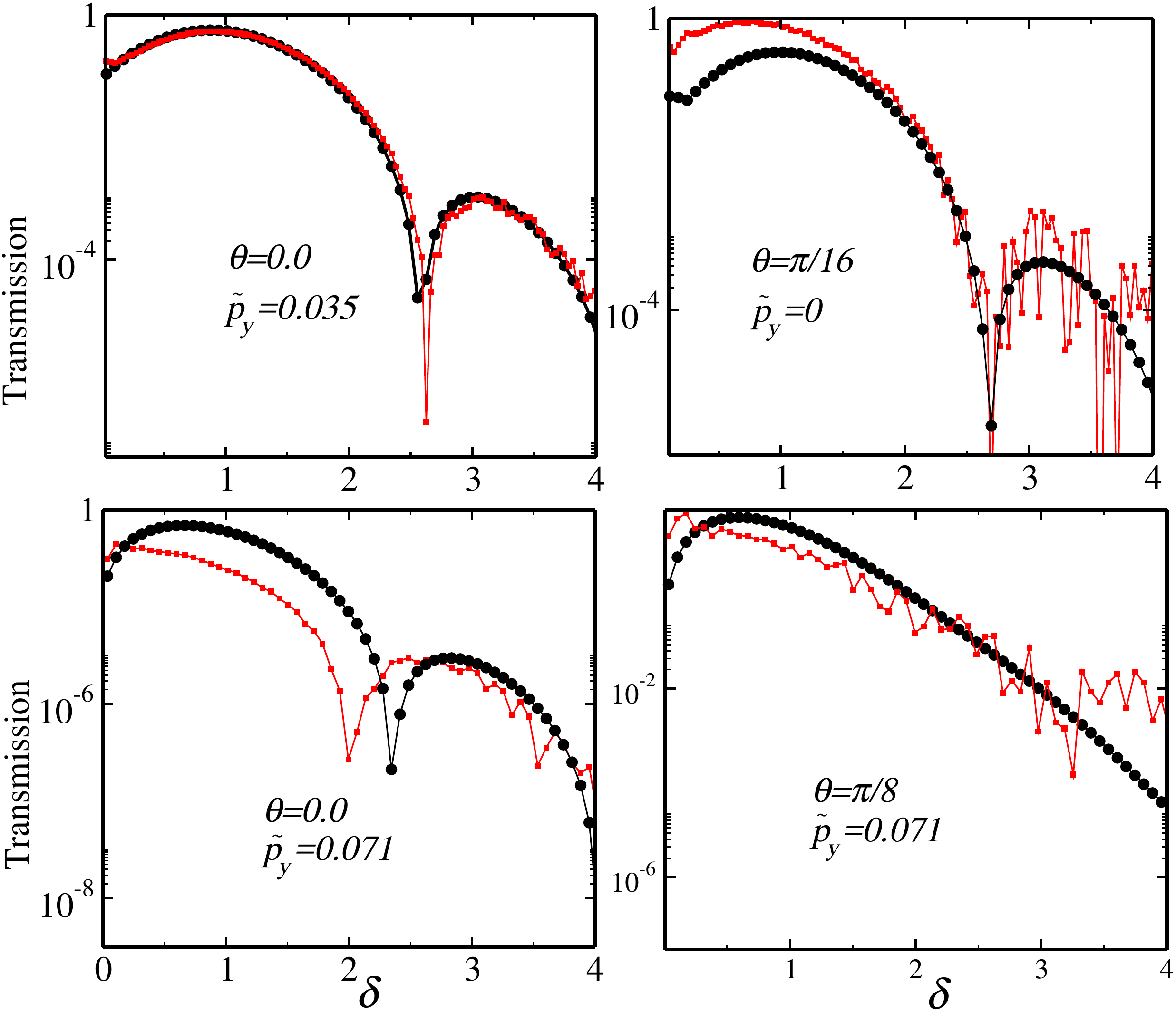}
\caption{ Comparison between WKB and numerical results for finite $\tilde p_y$ and $\theta$. For small $\tilde p_y$ there is good agreement  between these two for $\varphi\simeq 1.45$. However, for large values of $\theta$ and $\tilde p_y$ the numerical solutions become unstable. Parameters are taken same as in Fig.~(\ref{ky:trandel})}.
\label{wkb:numeric}
\end{figure}
We now solve these coupled differential equations within a suitably chosen range of $\tilde p_x^{\rm min}<\tilde p_x<\tilde p_x^{\rm max}$. We take $a(\tilde p_x^{\rm min})=1$ and $b(\tilde p_x^{\rm min})=0$ as boundary conditions for this problem. With this, the transmission is given by the coefficient $b$ evaluated at $\tilde p_x^{\rm max}$.  For our numerical calculation, we take $\tilde p_x^{\rm min/max}=\mp\equiv\Lambda$, with $\Lambda=20\sqrt{\delta\eta/2}$. The inset of Fig.~(2) in the main text shows numerical results for $\tilde p_y=0$. In Fig.~(\ref{wkb:numeric}), we present additional numerical results for finite $\tilde p_y$ and $\theta$, and compare with the WKB results. Although for small $\tilde p_y$ and small $\theta$, we obtain reasonable agreement between WKB and numerical results, we observe deviations for large $\tilde p_y$ and/or large $\theta$. We believe these are owing to instability of the numerical integration of the more complex situation where $p_x$ is the solution to a general quartic equation (rather than one that has only even powers, as in BLG).

\end{document}